\pdfoutput=1 

\documentclass[journal,twoside, 10pt]{IEEEtran}
\usepackage{cite}
\usepackage{threeparttable}
\usepackage{amsmath,amssymb,amsfonts}
\usepackage{enumitem}
\ifCLASSOPTIONcompsoc
\usepackage[caption=false,font=normalsize,labelfon
t=sf,textfont=sf]{subfig}
\else
\usepackage[caption=false,font=footnotesize]{subfig}
\fi
\usepackage{multirow}
\usepackage[super]{nth}
\usepackage{layouts}
\usepackage[table,x11names]{xcolor}

\usepackage{float}

\usepackage{siunitx}
\usepackage{acronym}
\acrodef{fe}[FE]{flexible electronics}

\usepackage{algorithmic}
\usepackage{graphicx}
\usepackage{textcomp}
\usepackage{xcolor}
\usepackage{pifont}
\usepackage{makecell}
\def\BibTeX{{\rm B\kern-.05em{\sc i\kern-.025em b}\kern-.08em
    T\kern-.1667em\lower.7ex\hbox{E}\kern-.125emX}}

\newcommand{\red}[1]{{\color{black}#1}}
\newcommand{\blue}[1]{{\color{black}#1}}

\usepackage{flushend}

\begin{document}
\bstctlcite{IEEEexample:BSTcontrol} 
\setlength{\abovedisplayskip}{2ex}
\setlength{\belowdisplayskip}{2ex}

\title{
{\vspace{-0.6cm}\small This article is accepted for publication in \textit{IEEE Embedded Systems Letters} (DOI: 10.1109/LES.2024.3447412). \\
}
\vspace{-0.8\baselineskip}
\rule{\textwidth}{0.4pt}
Reducing ADC Front-end Costs During Training of On-sensor Printed Multilayer Perceptrons
}

\author{Florentia Afentaki,
        Paula Carolina Lozano Duarte,
        Georgios Zervakis,
        Mehdi B. Tahoori,~\IEEEmembership{Fellow,~IEEE}
        
\thanks{Manuscript received June 2 2024; accepted July 2 2024. \textit{Corresponding Author: Florentia Afentaki (afentaki@ceid.upatras.gr)}}%
\thanks{F. Afentaki and G. Zervakis are with the Computer Engineering \& Informatics Dept., University of Patras, Patras 26504, Greece.}
\thanks{P.~C.~Lozano Duarte and M.~B.~Tahoori are with the Department of Computer Science, Karlsruhe Institute of Technology, Karlsruhe 76131, Germany.}%
\thanks{This work is partially by the European Research Council (ERC), and co-funded by the H.F.R.I call “Basic research Financing (Horizontal support of all Sciences)” under the National Recovery and Resilience Plan “Greece 2.0” (H.F.R.I. Project Number: 17048).}
}

\markboth{IEEE EMBEDDED SYSTEMS LETTERS}{F. Afentaki \MakeLowercase{\textit{et al.}}: Reducing ADC Front-end Costs During Training of On-sensor Printed Multilayer Perceptrons}

\maketitle

\begin{abstract}
Printed electronics technology offers a cost-effective and fully-customizable solution to computational needs beyond the capabilities of traditional silicon technologies, offering advantages such as on-demand manufacturing and conformal, low-cost hardware.
However, the low-resolution fabrication of printed electronics, which results in large feature sizes, poses a challenge for integrating complex designs like those of machine learning~(ML) classification systems.
Current literature optimizes only the Multilayer Perceptron~(MLP) circuit within the classification system, while the cost of analog-to-digital converters~(ADCs) is overlooked. 
Printed applications frequently require on-sensor processing, yet while the digital classifier has been extensively optimized, the analog-to-digital interfacing, specifically the ADCs, dominates the total area and energy consumption.
In this work, we target digital printed MLP classifiers and we propose the design of customized ADCs per MLP's input which involves minimizing the distinct represented numbers for each input, simplifying thus the ADC's circuitry.
Incorporating this ADC optimization in the MLP training, enables eliminating ADC levels and the respective comparators, while still maintaining high classification accuracy.
Our approach achieves 11.2x lower ADC area for less than 5\% accuracy drop across varying MLPs.

\end{abstract}


\section{Introduction }\label{sec:intro}

Recently, there has been a growing trend fueled by the fourth industrial revolution and the Internet of Things to integrate intelligence into everyday items.
Applications like wearables, fast-moving consumer goods, basic healthcare devices, and disposable sensors for pharmaceuticals have not yet, fully incorporate computing capabilities~\cite{Bleier:ISCA:2020:printedmicro}.
These products need computing technology that is ultra-low cost, thin, and conformal. 
Traditional lithography-based CMOS technologies can't meet these requirements, limiting computing's reach~\cite{Henkel:ICCAD2022:expedition}. 

Printed electronics (PE) offer a promising solution with on-demand, ultra-low cost fabrication, ideal for short-lifetime, disposable products. 
PE uses various printing methods like jet, screen, or gravure printing~\cite{Henkel:ICCAD2022:expedition}.
These techniques are mask-less, portable, and additive, reducing manufacturing costs and production times~\cite{Henkel:ICCAD2022:expedition}.
The simplicity of additive manufacturing allow ultra low-cost, at sub-cent levels, electronic circuits.
However, this low precision fabrication, results in higher device latency and lower integration density compared to silicon VLSI systems~\cite{Henkel:ICCAD2022:expedition}, making the design of more complex circuits a challenge in PE. 
Nevertheless, the target applications are viable in printed electronics due to their relaxed frequency and precision requirements~\cite{Henkel:ICCAD2022:expedition}.
We focus on Electrolyte-Gated FET~(EGFET) technology, which has low supply voltage ($\leq1\si{\volt}$), making it suitable for battery-powered applications~\cite{Bleier:ISCA:2020:printedmicro}.

\begin{figure}[!t]
\centering
\includegraphics[width=0.78\columnwidth]{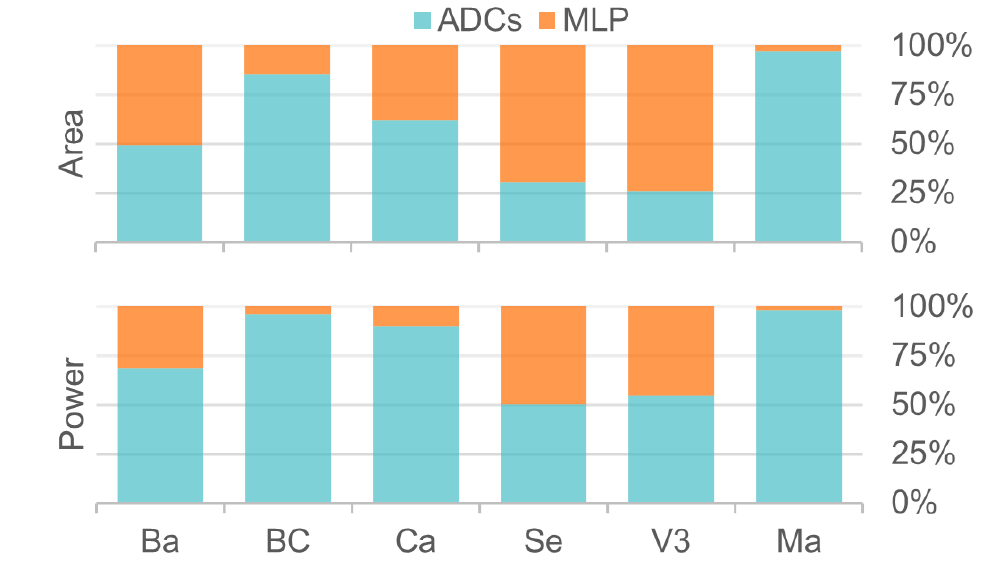}
\vspace{-3ex}
\caption{Area and Power Evaluation of the printed classification system in~\cite{Afentaki:ICCAD23:hollistic}.}
\vspace{-5ex}
\label{fig:motivation}
\end{figure}

PE's target applications require smart sensor processing, which starts with analog frontend to digitize analog sensor data using ADCs, followed by machine learning classifiers, such as Multilayer Perceptrons~(MLPs)~\cite{Henkel:ICCAD2022:expedition}. 
However the complexity of these classification system oppose a challenge to their realization in PE due to their large gate count. 
To mitigate this limitation, exploiting the high customization capabilities of the PE technology by bespoke implementations in which the hardware is tailored to a specific dataset and model were proposed~\cite{Mubarik:MICRO:2020:printedml}.
In~\cite{Afentaki:ICCAD23:hollistic, Afentaki:DATE2024:embedding, Armeniakos:TC2023:codesign, Kokkinis:DATE2023}, the authors combined the bespoke architecture alongside the well established Approximation Computing paradigm where a small, accuracy loss resulted in significant area and power gains of the MLP classifier.
Nevertheless, all the previous works focused only on the reduction of the MLP classifier inside the overall classification system, neglecting the area and power consumption of the Analog-to-Digital Converts~(ADCs).
In \figurename~\ref{fig:motivation}, an area and power analysis within the classification system is presented, by using the MLP in~\cite{Afentaki:ICCAD23:hollistic}.
As shown in~\figurename\ref{fig:motivation}, since the MLP classifier is optimized using approximate bespoke mapping, the ADC ratio in the classification system consuming on average 58\% area and 74\% power of the entire classification system, making the ADCs the dominant source of area and power overhead in the classification system.
Inspired by the fact that different sensor data have different distributions in a given range (e.g., 4bits), not all the representations are required and thus high accuracy can be achieved albeit discard them.
Leveraging the above, in this work, we propose the design of bespoke pruned ADCs for each input with the minimum possible representations required saving thus hardware by removing the circuitry of the unused input representation.
In more details, we use a Genetic Algorithm~(GA) to explore which representations of the ADC can be pruned alongside a quantization-aware training~(QAT) of the MLP.
Further, our ADC-optimization is orthogonal to any other training approach.
\blue{
In the literature some ADC optimization have been proposed in other technologies rather than in printed electronics.
In~\cite{song:HPCA2017:Pipelayer}, a spike-based scheme avoids ADCs by a comparator-register architecture but both of these components are hardware expensive in printed electronics\cite{Mubarik:MICRO:2020:printedml}. 
In~\cite{ogbogu:TCAD2022:acceleratingCrossbar}, prune crossbars to eliminate ADCs in ReRAM architectures, which in our architectures is equivalent to feature reduction. Our approach goes further by optimizing the remaining ADCs.
}

To the best of our knowledge, this is the first time that such a framework\footnote{{\scriptsize https://github.com/floAfentaki/Approximation-Techniques-Targeting-Printed-MLPs}} is proposed for ADC-efficient printed MLP-based classification systems.
Our experiments across various datasets showcase that our framework reduces both area and power of the required ADCs on average \red{$11.2\times$} and \red{$13.2\times$} respectively.
\section{Proposed ADC-aware Methodology }\label{sec:framework}

\begin{figure}[!t]
\centering
\includegraphics[width=\columnwidth]{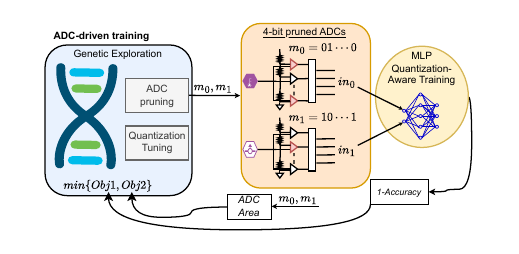}
\vspace{-5.5ex}
\caption{A high-level overview of the proposed ADC-aware methodology. }
\vspace{-3ex}
\label{fig:framework}
\end{figure}

This section describes the proposed ADC-aware methodology, which is illustrated in \figurename~\ref{fig:framework}.
The target ADC in our framework is a flash ADC, which traditionally generates $2^N$ quantization levels for an N-bit ADC. 
However, our area-reduction strategy involves using partial ADCs that produce only $k$ quantization levels, where $k<2^N$. 
This approach is grounded in the idea that not all quantization levels are utilized in the output of the ADC. 
By integrating a training algorithm that minimizes the number of levels while maximizing classification accuracy, we reduce ADC costs in terms of both area and power.
To implement this, we use a proxy area model for the partial ADC, that considers the costs of comparators and encoders. 
This area model serves as a secondary objective in our multi-objective learning framework, alongside the primary accuracy loss, guiding the optimization process to achieve efficient and effective classification systems.
The best area-efficient ADCs are searched and explored through a genetic algorithm, which enables the identification of optimal configurations that balance area and accuracy.

\subsection{ADC Pruning}\label{sec:ADCs}

ADCs are a necessity in a classification system for digitizing the analog sensor data to be processed by digital classifier hardware.
In this work, we focus on Flash ADCs for their simple architecture, suitable for low-precision printed applications. 
Flash ADCs provide rapid data conversion, ensuring timely data availability during short energy windows. 
Despite high power use during conversion, their low-power operation overall reduces average power and energy consumption, fitting well with energy-harvesting systems.

\begin{figure}[!t]
\centering
\includegraphics[width=\columnwidth]{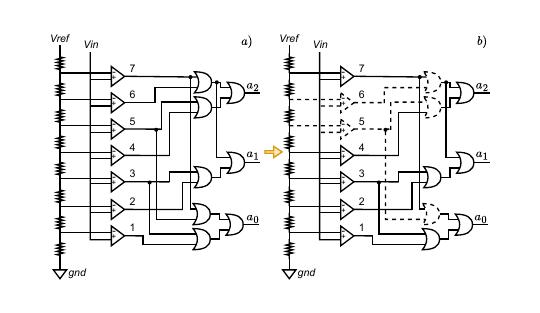}
\vspace{-7ex}
\caption{Schematic of: a) conventional 3-bit Flash ADC and b) an example of an equivalent bespoke pruned ADC}
\vspace{-3ex}
\label{fig:ADC}
\end{figure}

\figurename\ref{fig:ADC}a depicts the architecture of a 3-bit flash ADC.
In general, for an N-bit ADC, the analog input voltage~(Vin) is compared among \(2^N-1\) comparators with \(2^N-1\) different levels. 
The number of these different levels corresponds to the resolution of the ADC.
The different levels are generated by using a reference voltage~(Vref) which is connected to a resistance ladder in order to divide the Vref into \(2^N\) equally levels in the range \([0, Vref]\).
When an analog input voltage Vin is applied to the ADC, each comparator compares Vin with its corresponding  Vref. 
If Vin is larger than Vref, the comparator outputs a high signal (1); otherwise, it outputs a low signal (0), creating a digital thermometer code representation.
The translation of this thermometer code to its binary counterpart i.e., \((0011 1111)_{TC} \rightarrow {110}_{2} \) is delivered by a digital priority encoder which is implemented by multiple OR. 

Leveraging the low non-recurring costs of the PE technology, we are able to design bespoke ADCs for each of the sensor inputs of our classifier.
This means that each ADC is specifically designed for the respective sensor values allowing us to keep only the necessary representations for each sensor input that represent best each sensor distribution.
\figurename\ref{fig:ADC}b depicts the an example of the architecture of a bespoke pruned 3-bit flash ADC.
In this example the levels \textit{5} and \textit{6} are pruned from the representations of the ADC.
As can been seen, pruning the some of the levels of the ADC, does not only removes the equivalent comparators but also reduces the complexity of priority encoder by simplifying the OR-logic required (OR by zero is identity).
When an analog input Voltage \(V_{in}\) is applied to the bespoke pruned ADC, each of the remaining comparators compares \(V_{in}\) with the corresponding \(V_{ref}\).
Due to the pruned \(V_{ref}\) levels, the inference of the analog \(V_{in}\) leads to a restrained subset of representations in thermometer code and respectively to its binary counterpart.
For example, if the sensor data input by using the conventional ADC presented into \figurename\ref{fig:ADC}a was denoted into the values \((0011 1111)_{TC} \rightarrow {110}_{2}\), then by using the bespoke pruned ADC in the \figurename\ref{fig:ADC}b the sensor data will be denoted into the values \((0000 1111)_{TC} \rightarrow {100}_{2}\), since the corresponding level \textit{6} is removed alongside the follow level \textit{5} leading the sensor input to be digitized as the right after smaller representation i.e, \({110}_{2}\).
It is important to note, that pruning the ADC is different than simply selecting a lower bitwidth ADC e.g., $ 3\text{-bit ADC} \rightarrow 2\text{-bit ADC}$, since our approach finds the subset from all the levels in any given bitwidth.
For example in \figurename~\ref{fig:ADC} the bespoke pruned 3-bit ADC has $k=6$ quantization levels rather than the conventional $2^3=8$ quantization levels. 
  
\subsection{ADC-Area Model}

In order to capture the ADC area reduction during the exploration process we developed a Python script which calculates a proxy area model of the pruned bespoke ADC based on the remaining representation level after the ADC pruning.
\blue{
Using a Python script for a proxy area model allows the parallel evaluation, be independently of EDA license constraints.
}
Specifically, the flash ADC can be divided into the three distinct parts; the resistance ladder, the comparators and the encoder.
Out of the three of them the resistance ladder is not affected by pruning the levels of representation since we keep the uniform relation between them.
However as can be seen in \figurename~\ref{fig:ADC} our pruning procedure reduces the number of the comparators. 
The number of comparators in the final pruned ADC is equal to the number of the remaining levels.
As for the encoder reduction we use a simple procedure to calculate the number of OR gates required based on the pruned subset of the representations. 
Specifically, inside a generic \(N\)-bit ADC, each output of the generic encoder \(a_i, \text{ where } i\in[0,N-1]\), is a bitwise OR between \(2^N/2\) pre-determined levels.
When a representation is pruned, the corresponding representation is removed from these levels for each output \(a_i\).
Thus, our script tracks the levels that are used in each of the output bit \(a_i\) of the encoder and can easily calculate the number of required ORs and subsequently the area of the bespoke pruned ADC.
The correctness of our area model is verified by the almost perfect correlation score \red{\(0.95\)} between all the possible \red{\(2^{15}\)} samples between our model and real synthesis area values.

\subsection{ADC-driven Training Flow}

As depicted in \figurename~\ref{fig:framework}, our ADC-driven training flow utilize a framework implements an ADC-aware exploration targeting to find the best bespoke pruned ADCs in terms of the classifier accuracy and ADC area.
Thus a multi-objective minimization optimization problem is formed with constraints the accuracy loss of the classifier and the overall ADC area.
Due to simplicity, low computational complexity, and enhanced convergence, we employ the Non-dominated Sorting Genetic Algorithm II (NSGA-II~\cite{Deb:NSGA:2002}) to address this multi-objective problem.

It's important to note that pruning the ADC and removing representation naturally leads to diminished accuracy from the classifier perspective, especially if the remaining levels do not capture the sensor-based input's distribution.
Our approach searches for the ideal bespoke ADCs that follows each of the sensor inputs' distribution. 
That means that the accuracy degradation is minimal while the ADC area gains are optimal.
In the case of a 4-bit ADC, our ADC-pruning procedure explores the minimum number of levels $k$ from the complete set of 16-uniform levels of the conventional ADC, for each input sensor data. 
In order to encode the ADC-pruning procedure we adopt a set masking parameters \(m_i={0,1}\) one for each of 16-uniform different levels for each sensor data, which for \(m_i=0 \text{ and } m_i=1\) removes or keep the quantization level respectively. 
Additionally, in order to further enhance the balance between the area and accuracy, the quantization precision of the coefficients, the activation function and the tuning parameter i.e., batch size and number of epochs, of the QAT are explored alongside the ADC-pruning within the GA.
Thus, the chromosome is a set of the aforementioned parameters i.e., the masks $m_i$ for each of the $2^N$ levels for each input sensor, the quantization and tuning parameters of the QAT.
\figurename~\ref{fig:framework} illustrates the ADC-aware flow, that uses two sensor inputs with two bespoke pruned ADCs which are further connected as the inputs of the MLP QAT.
As shown, in each generation, the GA by using mutation and crossover, searches the optimal parameters which prune each of the bespoke pruned ADC for each sensor input while configure the best the QAT training of the MLP classifier.
The evaluation of each chromosome is conducted by the fast-non-dominated-sort algorithm using as objective the accuracy miss of the QAT and the overall area of the partial ADCs.
\section{Results and Evaluation}\label{sec:results}

\subsection{Experimental Setup}

We examine \red{six} datasets, namely Balance~(Ba), Breast Cancer~(BC), Cardiotocography~(Ca), Mammographic~(Ma), Seeds~(Se), and Vertebral Column 3~(V3), from~\cite{Dua:2019:uci}. 
These datasets could form realistic printed applications as they utilize sensor inputs suitable for printed applications and have low precision, duty cycle, and sample rate requirements~\cite{Bleier:ISCA:2020:printedmicro}.

Both the MLP architecture and the datasets used in this work follow the~\cite{Mubarik:MICRO:2020:printedml, Armeniakos:TC2023:codesign, Kokkinis:DATE2023, Afentaki:ICCAD23:hollistic, Afentaki:DATE2024:embedding} ensuring a fair comparison for our analysis.
The inputs are normalized to $[0, 1]$ as in~\cite{Mubarik:MICRO:2020:printedml, Armeniakos:TC2023:codesign, Afentaki:ICCAD23:hollistic, Afentaki:DATE2024:embedding, Kokkinis:DATE2023} and are randomly stratified split into $70\%$ for the train and $30\%$ for  the test set.
It is important to note that in this work we do not consider any approximation for our MLP classifier beside precision approximation i.e., power-of-2 quantization, such as accumulation approximation like other literature consider in~\cite{Armeniakos:TC2023:codesign, Afentaki:DATE2024:embedding, Afentaki:ICCAD23:hollistic}.
Thus, our baseline are the bespoke power-of-2 printed MLPs circuits, designed following the approach outlined in~\cite{Kokkinis:DATE2023}, using $8$-bit power-of-2 fixed point weights and $4$-bit inputs.
Clock period of $200\si{\milli\second}$ are applied to all MLPs.
Such delay values align with typical PE performance~\cite{Afentaki:DATE2024:embedding}.
Mutation and crossover operators of the GA are set to $0.2\%$ \text{and} $0.7\%$ respectively.
For the quantization-aware training the QKeras framework was used, an extension of Keras by Google~\cite{coelho:Qkeras}.
All circuits are synthesized using Synopsys Design Compiler S-2021.06 and mapped to the printed EGFET library~\cite{Bleier:ISCA:2020:printedmicro}, while VCS T-2022.06 and PrimeTime T-2022.03 are used for simulation and power analysis.
The accuracy is reported on the test dataset, and all designs are synthesized at a relaxed clock period to improve area efficiency.

\begin{figure}[!t]
\centering
\includegraphics[width=\columnwidth]{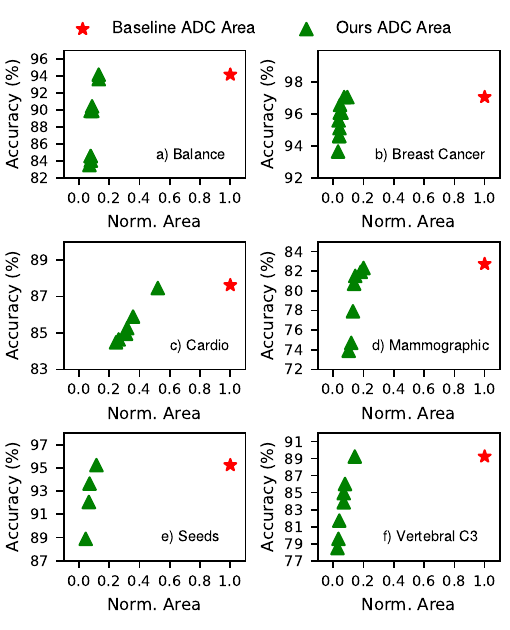}
\vspace{-6ex}
\caption{Accuracy vs normalized area Pareto space of the ADCs.}
\vspace{-3ex}
\label{fig:pareto_front}
\end{figure}

\subsection{Evaluation against the State-Of-The-Art}

First, we assess the effectiveness of our framework in the ADC reduction.
\figurename~\ref{fig:pareto_front}, illustrates the bespoke ADC area over accuracy reduction after the proposed pruning methodology is applied.
The area values are obtained by our area model while the accuracy is referring to the classification accuracy of the classifier of the specific ADCs.
The area value is normalized w.r.t. the area of the corresponding conventional ADCs as all works before the proposed one used.
Compared to the conventional ADCs, our ADCs pruning achieves \red{\(11.2\times\)} area and \red{\(13.2\times\)} power reduction on average for less than $5$\% lower accuracy.
For up to 5\% accuracy degradation, the area gains range from \red{\(3.3\times\)} for \red{Cardio} and go up to \red{\(15\times\)} for \red{Seeds}.

In Table~\ref{tab:res}, we evaluate the impact of our work at classification level. 
To that end, we compare against the area-efficient state-of-the-art MLP design in~~\cite{Kokkinis:DATE2023}, which uses only pow2 for weights (i.e., no costly multipliers).
From all the designs of the~\figurename\ref{fig:pareto_front}, we select the ones with up to \(1\%\) accuracy loss compare to the baseline~\cite{Kokkinis:DATE2023}.
As shown in Table~\ref{tab:res}, our methodology results in a smart sensor-based system that significantly outperforms the state-of-the-art~\cite{Kokkinis:DATE2023}.
Specifically, compared to~\cite{Kokkinis:DATE2023} and account for the costs of ADCs, our classification achieve on average \red{$2\times$} area reduction and  \red{$6.9\times$} power reduction, while the gains range from \red{\(1.3\times\)} up to \red{\(2.4\times\)} in  area and from \red{\(3.3\times\)} up to \red{\(9.7\times\)} power respectively.

Our ADC-aware training method, executed on an AMD EPYC 7552 with 256GB RAM, averages only \red{120} minutes.
Even with the increased parameter count due to ADC pruning for each input sensor, the runtime impact is minimal, comparable to hardware-unaware conventional training. 
Finally, the proposed ADC pruning methodology can be implemented on top of any MLP approximations since it only affects the ADCs i.e., the inputs of the MLP, thus can leverage any other digital optimizations e.g., addition approximation~\cite{Afentaki:DATE2024:embedding, Afentaki:ICCAD23:hollistic}.
\begin{table}[t!]\label{tab:res}
\setlength\tabcolsep{2pt}
\caption{Evaluation of our printed classification system using MLPs for up to 1\%. 
}
\vspace{-1ex}
\footnotesize
\centering
\renewcommand{\arraystretch}{1.1}
\begin{threeparttable}
\begin{tabular}{|l|cc|cc|cc|cc|cc|cc|cc|}
\cline{1-15}
\multicolumn{1}{|c|}{\multirow{3}{*}{\textbf{Data}}} &  
\multicolumn{6}{c|}{\textbf{~\cite{Kokkinis:DATE2023}}} &  \multicolumn{6}{c|}{\textbf{Ours}} 
&\multirow{3}{*}{\begin{tabular}{@{}c@{}}Area \\ Gains\end{tabular}}
&\multirow{3}{*}{\begin{tabular}{@{}c@{}}Power \\ Gains\end{tabular}}\\ 
\cline{2-13}

&  
\multicolumn{2}{c|}{\footnotesize{\textbf{ADCs}}} & \multicolumn{2}{c|}{\footnotesize{\textbf{MLP}}} & \multicolumn{2}{c|}{\footnotesize{\textbf{Total}}}&
\multicolumn{2}{c|}{\footnotesize{\textbf{ADCs}}} & \multicolumn{2}{c|}{\footnotesize{\textbf{MLP}}} & \multicolumn{2}{c|}{\footnotesize{\textbf{Total}}} & \multicolumn{2}{c|}{\textbf{}}\\

\multicolumn{1}{|c|}{\textbf{}}
&  
\textbf{A\tnote{1}}&\textbf{P\tnote{2}}&
\textbf{A\tnote{1}}&\textbf{P\tnote{2}}&
\textbf{A\tnote{1}}&\textbf{P\tnote{2}}&
\textbf{A\tnote{1}}&\textbf{P\tnote{2}}&
\textbf{A\tnote{1}}&\textbf{P\tnote{2}}&
\textbf{A\tnote{1}}&\textbf{P\tnote{2}}& 
\multicolumn{2}{c|}{\textbf{}}
\\ 
\cline{14-15}
\cline{1-13}

\textbf{Ba} & 0.7 & 5.2 & 0.5 & 1.2 & 1.2 & 6.4 & 0.1 & 0.9 & 0.4 & 0.1 & 0.5 & 1 &\textbf{2.4$\times$} &\textbf{6.4$\times$}\\
\textbf{BC} & 1.5 & 12 & 5 & 17 & 5.7 & 29 & 0.1 & 2.5 & 3.3 & 1.2 & 3.4 & 3.7 &\textbf{1.7$\times$} &\textbf{7.8$\times$}\\
\textbf{Ca} & 3.6 & 27 & 9 & 34 & 12.6 & 61 & 1.9 & 16 & 7.5 & 2.5 & 9.4 & 18.5 &\textbf{1.3$\times$}&\textbf{3.3$\times$}\\
\textbf{Ma} & 0.9 & 6.5 & 0.5 & 1.8 & 1.4& 8.3 & 0.1 & 1.3 & 0.7 & 0.3 & 0.8 & 1.6 &\textbf{1.8$\times$}&\textbf{5.2$\times$}\\
\textbf{Se} & 1.2 & 9 & 4.5 & 20 & 5.7&  29  & 0.2 & 1.9 & 2.3 & 1.1 & 2.5 & 3 &\textbf{2.3$\times$}&\textbf{9.7$\times$}\\
\textbf{V3} & 1.0 & 7.8 & 5.2 &  17 & 6.2 & 25 & 0.1 & 1.7 & 2.6 & 1 & 2.7 & 2.7 &\textbf{2.3$\times$}&\textbf{9.2$\times$}\\ \hline

\end{tabular}

\begin{tablenotes}\footnotesize
\item[] 
$^1$Area in \(\si{\centi\meter}^2\). 
$^2$Power in \(\si{\milli\watt}\). 

\vspace{-3.5ex}
\end{tablenotes}
\end{threeparttable}
\end{table}
\section{Conclusion}

Printed electronics is a promising complementary solution to silicon counterparts for on-sensor processing in cheap wearables, disposables, medical devices and consumer market. 
While the main research focus is on reducing the cost of digital classifiers using approximate bespoke hardware mapping, the Analog Digital Converters (ADCs) at the analog sensory inputs dominate the total system costs in terms of area and power.
In this work, we propose an ADC pruning methodology that efficiently integrates the training of the MLP classifier with eliminating quantization levels of ADC, successfully balancing the classification accuracy of the MLP and ADC costs.
Overall, our approach is seamlessly applied and achieves higher ADC-cost savings, demonstrating superior hardware efficiency and high accuracy compared to the state of the art.

\bibliographystyle{IEEEtran}
\bibliography{IEEEabrv,references}

\end{document}